\documentclass[sigconf, 9pt]{acmart}

\pdfoutput=1

\renewcommand\footnotetextcopyrightpermission[1]{} % removes footnote with conference information in first column
\makeatletter
\renewcommand\@formatdoi[1]{\ignorespaces}
\makeatother

\acmConference[IoT S\&P'19]{Workshop on the Internet of Things Security and Privacy}{London, UK}

\usepackage[utf8]{inputenc}
\usepackage{amsmath}
\usepackage{amsfonts}
\usepackage{amsthm}

\usepackage{float}

\usepackage{graphicx}
\graphicspath{{Figures/}}
\usepackage{epstopdf}

\usepackage{subcaption} % for acm format
\usepackage{epsfig}
\usepackage{url}
\usepackage{comment}
\usepackage{graphics}
\usepackage{soul} 
\usepackage{multirow}
\usepackage[export]{adjustbox}

\usepackage{todonotes} % to do (added by monowar)

\usepackage{xfrac}

\usepackage{graphicx, array, blindtext} % for figure inside table

\usepackage{graphicx}

\usepackage{wrapfig}

\usepackage{algorithm}  
\usepackage{algorithmic}

\usepackage{listings}
\usepackage{psfrag,wrapfig}

\usepackage{enumerate}% http://ctan.org/pkg/enumerate
\usepackage{paralist} %inline list
\usepackage{subfiles} % for multifile easy compilation

\usepackage{array}
\usepackage{ragged2e}
\newcolumntype{P}[1]{>{\RaggedRight\hspace{0pt}}p{#1}}

\usepackage{multirow}

\usepackage{nimbusmononarrow}
\usepackage{lstlinebgrd}

\newcommand{\eg}{{\it e.g.,}\xspace}
\newcommand{\viz}{{\it viz.,}\xspace}

\newcommand{\ie}{{\it i.e.,}\xspace}
\newcommand{\etc}{{\it etc.}}
\newcommand{\ci}{{\it (i) }}
\newcommand{\cii}{{\it (ii) }}
\newcommand{\ciii}{{\it (iii) }}
\newcommand{\civ}{{\it (iv) }}
\newcommand{\ca}{{\it (a) }}
\newcommand{\cb}{{\it (b) }}
\newcommand{\cc}{{\it (c) }}

\theoremstyle{acmdefinition}

\newcommand{\pnametz}{\xspace{Contego-TEE}\xspace}
\newcommand{\nw}{\xspace{NW}\xspace}
\newcommand{\sw}{\xspace{SW}\xspace}
\newcommand{\tzignore}{\xspace{IGNORE}\xspace}
\newcommand{\tzfailsafe}{\xspace{FAIL-SAFE}\xspace}

\begin{document}

% \title{Securing Real-Time Actuators using Trusted Execution Environment}
\title{Protecting Actuators in Safety-Critical IoT Systems from Control Spoofing Attacks}

% Sibin't tittle suggestion [(1) for main conference]
% 1. Securing Real-Time IoT Systems from Control Spoofing Attacks
% 2. Using Trusted Execution Environments to Prevent Control Spoofing Attacks in Safety-Critical IoT Systems
% 3. Protecting Actuators in Safety-Critical IoT Systems from Control Spoofing Attacks

%
% The "author" command and its associated commands are used to define
% the authors and their affiliations.
% Of note is the shared affiliation of the first two authors, and the
% "authornote" and "authornotemark" commands
% used to denote shared contribution to the research.
 \author{Monowar Hasan}
 \affiliation{\institution{Dept. of Computer Science, University of Illinois}}
 \email{mhasan11@illinois.edu}

 \author{Sibin Mohan}
 \affiliation{\institution{Dept. of Computer Science, University of Illinois}}
 \email{sibin@illinois.edu}

%
%\author{Monowar Hasan, Sibin Mohan}
%\affiliation{\institution{Dept. of Computer Science, University of Illinois} \{mhasan11, sibin\} @illinois.edu}
%% \email{  \{ mhasan11, sibin \} @illinois.edu}

\renewcommand{\shortauthors}{Hasan and Mohan, et al.}

\begin{abstract}
 In this paper, we propose a framework called \pnametz to secure Internet-of-Things (IoT) edge devices with timing requirements from control spoofing attacks where an adversary sends malicious control signals to the actuators. We use a trusted computing base available in commodity processors (such as ARM TrustZone) and propose an invariant checking mechanism to ensure the security and safety of the physical system. A working prototype of \pnametz was developed using embedded Linux kernel. We demonstrate the feasibility of our approach for a robotic vehicle running on an ARM-based platform.
\end{abstract}

\maketitle

% paper content
\section{Introduction}

Today's embedded and cyber-physical systems are
ubiquitous. A large number of critical cyber-physical systems (\eg autonomous cars, drones, manufacturing systems, power
grids, industrial control systems, \etc)
have real-time (RT) properties (\eg strict timing
and safety requirements).
The current trend is to connect embedded RT devices
to the Internet (\eg remote surveillance over wired/wireless network, connected vehicles through cellular wireless networks, \etc) and this gives rise to the real-time Internet-of-Things (RT-IoT)~\cite{mhasan_rtiot_sensors19}. RT-IoT systems are intended to provide better user experience through
stronger connectivity and better use of next-generation embedded
devices, albeit with safety-critical properties. RT-IoT systems are also increasingly becoming targets for cyber-attacks. A number of high-profile attacks on RT-IoT systems, \eg denial-of-service (DoS) attacks mounted
from IoT devices~\cite{ddos_iot_camera}, Stuxnet~\cite{stuxnet}, attack demonstrations by researchers on medical devices~\cite{security_medical} and automobiles~\cite{checkoway2011comprehensive} have shown that the threat is real.
Successful cyber attacks against such systems could lead to
problems more serious than just loss of data or availability
because of their critical nature~\cite{mhasan_rtiot_sensors19}.
Enabling security in RT-IoT, however, is often more challenging than
generic IoT due to additional timing/safety constraints imposed
by RT-enabled systems.

\begin{figure}[t]
%\begin{wrapfigure}{r}{0.5\columnwidth}
    %\hspace*{-2em}
    \centering
    \includegraphics[scale=0.345]{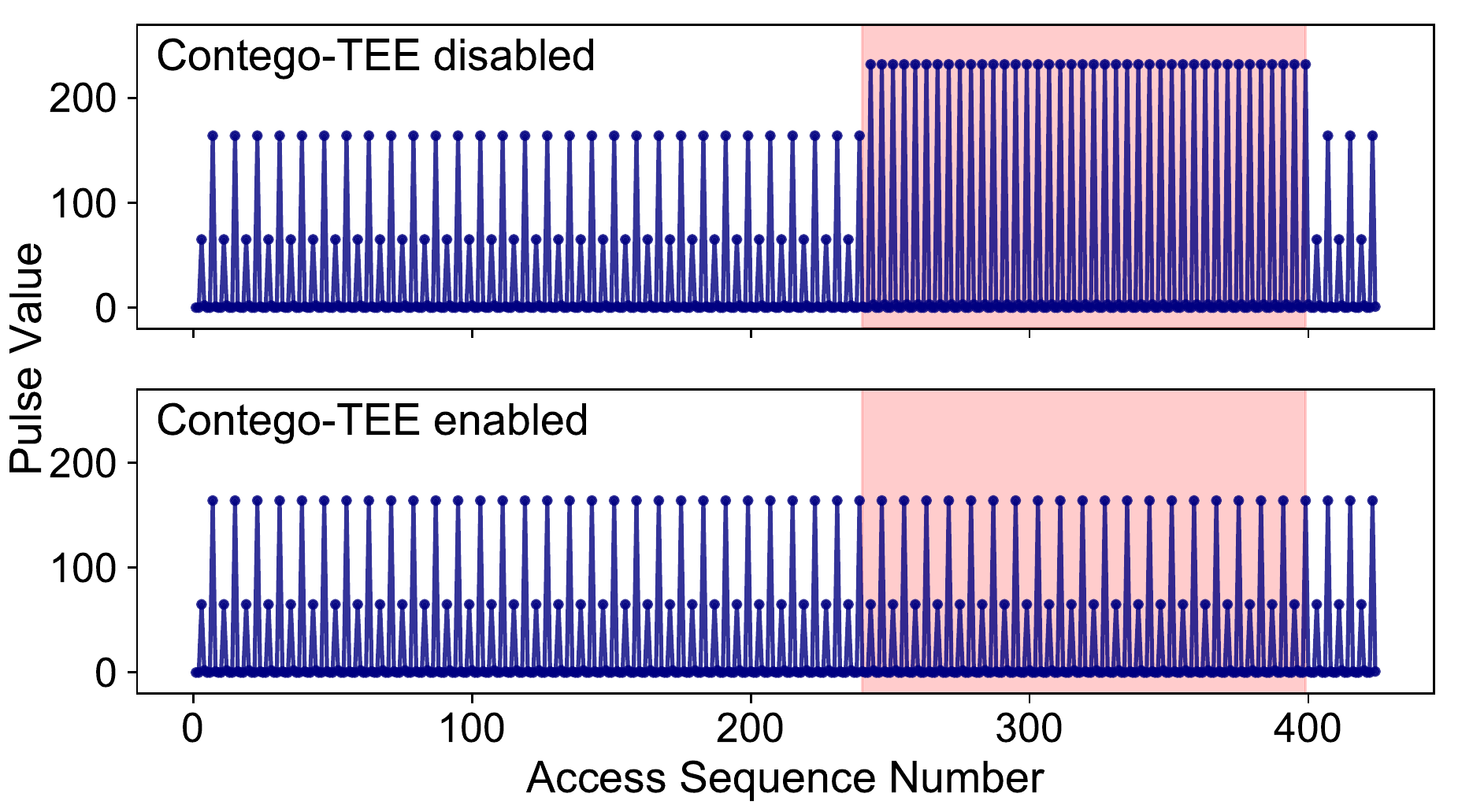}
    \caption{Demonstration of a control spoofing attack on a robotic control arm running embedded Linux.} 
    \label{fig:robot_arm_demo}
%\end{wrapfigure}
\end{figure}

Since RT-IoT systems are largely based on sensing and actuation, any false/spoofed command to the actuators can disrupt the normal operation of the physical plant. 
Commonly used open-source RT-IoT development stacks (such as Linux) do not provide explicit control over actuation signals. For instance, if the application task obtains permission (say, root or other privileged user access) to the peripheral interface (\eg I2C~\cite{i2c}), it is possible to send arbitrary signals to the actuators.  Let us consider an industrial robotic arm (running an embedded variant of Linux in an ARM Cortex-A53 platform~\cite{rpi3}) that periodically opens and closes the grip to drop off and pick up objects in an assembly line. The movement of the grip is controlled by a servo. We use an open-source implementation~\cite{robot_arm_src} for this robotic arm where each operation is represented by a pulse value $\mathtt{x}$ (where $\mathtt{x}=577$ for \textsf{grip\_open()} and $\mathtt{x}=420$ for \textsf{grip\_close()}) and each pulse sends the following four $1$ byte command sequences to the servo registers: $\mathtt{0~\&~0xFF}$, $\mathtt{0>>8}$, $\mathtt{x~\&~0xFF}$, $\mathtt{x>>8}$. An example of a spoofing attack for this control arm  is presented in Fig.~\ref{fig:robot_arm_demo} (x-axis is the servo access sequence number and y-axis is the corresponding pulse value). Without any actuation command validation, it is possible to send arbitrary (high) pulses to the servo registers that prevents the grip from picking up/dropping objects (showing in the shaded region, see the top figure) that is not otherwise possible when our scheme (called \pnametz, see \S\ref{sec:act_mon_frm} for details) is enabled (bottom figure).

Our proposed framework,  \pnametz, prevents the sending of malicious/undesired commands to physical actuators and ensures safety of the system. Specifically, we use the concept of Trusted Execution Environments (TEEs)~\cite{7345265_tee} available in commodity processors (\eg ARM TrustZone~\cite{trustzone_survey}, Intel SGX~\cite{intel_sgx}) to ensure that our protection mechanisms can not be disabled even if the host OS is compromised. We develop a rule-based invariant checking and access control mechanism as well as design-time (schedulability) tests to ensure timing and safety requirements of the system. \pnametz specifically designed for \textit{legacy systems} developed with Commodity-Off-The-Shelf (COTS) components and \textit{does not require any modification to the application code/logic}.

In this paper we present the following contributions.

\begin{itemize}
    \item A new framework called \pnametz to secure COTS-based RT-IoT systems against attacks that spoof control signals (\S\ref{sec:design}). 
    
    \item A runtime, rule-based invariant checking mechanism as well as design-time analysis to ensure security (and safety) of the physical plant (\S\ref{sec:inv_n_timing}).
    
    \item An open source implementation and patch to the (embedded) Linux kernel that includes the \pnametz functionality (\S\ref{sec:implementaion}).
\end{itemize}

We use ARM TrustZone as a TEE and implemement our solution in an  ARM Cortex-A53 board (\ie Raspberry Pi~\cite{rpi3}). We also demonstrate the viability of our approach using a COTS rover platform (\S\ref{sec:rov_case_study}). 
\section{System and Adversary Model}

In the following, we first present background on RT-IoT systems and give an overview of a TEE-based architecture (\eg ARM TrustZone).
%in \S\ref{sec:background}. 
We then introduce our system model (\S\ref{sec:sys_model}) and describe our assumptions on the adversarial capabilities (\S\ref{sec:adv_model}).

\subsection{Preliminaries} \label{sec:background}

\subsubsection{RT-IoT Systems:}

RT-IoT systems comprise IoT edge devices with RT capabilities. RT systems are those that, apart from a requirement for functional correctness,
require that temporal properties be met as well. These temporal properties are often
presented in the form of \textit{deadlines}. The usefulness of results (say the performance of the actuators) produced by the system
drops after the passage of a deadline.
Some of the common properties and assumptions related to RT systems include~\cite{mhasan_rtiot_sensors19}: \ci periodic/sporadic execution of set of tasks\footnote{The `task' in RT-IoT systems can trivially be mapped with the concept of \textit{process} or \textit{thread} in general-purpose OSes.}, \cii strict timing and safety requirements, \ciii well-characterized execution time (\eg execution times in the worst-case are known for all loops), \civ limited resources (\eg memory, processing power and energy).

RT-IoT systems are often designed based on the periodic task model~\cite{Liu_n_Layland1973}, \ie each task $\tau_i$ is characterized by a tuple: $(C_i, T_i, D_i)$ where $C_i$ is the Worst Case Execution Time (WCET), $T_i$ is the period (\eg inter-invocation time) and $D_i$ is the deadline.
Schedulability tests~\cite{res_time_rts, bini2004schedulability} are used to determine
if all tasks in the system meet their respective deadlines. If they do, then the taskset is deemed to be \textit{`schedulable'} and the system is considered \textit{safe}.

\subsubsection{TEE and ARM TrustZone:}

TEE is a set of hardware and software-based security extensions where the processors maintain a separated
subsystem in addition to the traditional
OS components. TEE technology has been implemented
on commercial secure hardware such as ARM TrustZone~\cite{trustzone_survey}
and Intel SGX~\cite{intel_sgx}. In this work we consider TrustZone as the building block of \pnametz due to wide acceptability of ARM processors for embedded IoT systems -- although our framework can be ported into other TEE platforms without loss of generality. ARM partnered with GlobalPlatform and has defined
new TEE APIs~\cite{globalplatform_tee_api}. TrustZone encompasses the following major features~\cite{liu2018alidrone}: \ca  safe and secure boot (to ensure all software components are in a trusted state before launching the
OS); \cb isolated execution of critical applications (\ie in a secure enclave) and \cc protection for trusted applications data (in terms of integrity
and confidentiality).

ARM
TrustZone contains two different privilege
blocks: \ci \textit{Normal World} (\nw) and
\cii \textit{Secure World} (\sw). The \nw is the untrusted environment running a commodity untrusted OS where \sw is a protected computing block that only
runs privileged instructions.  \sw in TrustZone defines the memory regions
that can only be accessed by privileged instructions and the code that runs in the \sw has higher privilege than the \nw. Hardware logic ensures that the resources
in the secure world can not be accessed from the normal world (\eg if the code running in the \nw tries to
access protected memory regions, TrustZone throws a hardware
exception).
 The \sw instructions are 
triggered when a specific flag in the processor \eg Non-secure (NS) bit in the Secure Configuration Register (SCR) is not set. 
 These two worlds bridge via a software module referred to as \textit{Secure Monitor}.
%A \textit{Secure Monitor} is used to bridge the \nw and \sw.  
The context switch between the \nw and \sw is performed through a \textit{Secure Monitor Call} (SMC).

In this work we use the TrustZone functionality to prevent the malicious commands from being sent to the actuators (See \S\ref{sec:act_mon_frm} for details). We now present our system and adversary model.

\subsection{System Model} \label{sec:sys_model}

\begin{figure}[t]
    \centering
    \vspace{-1em}
    \includegraphics[scale=0.45]{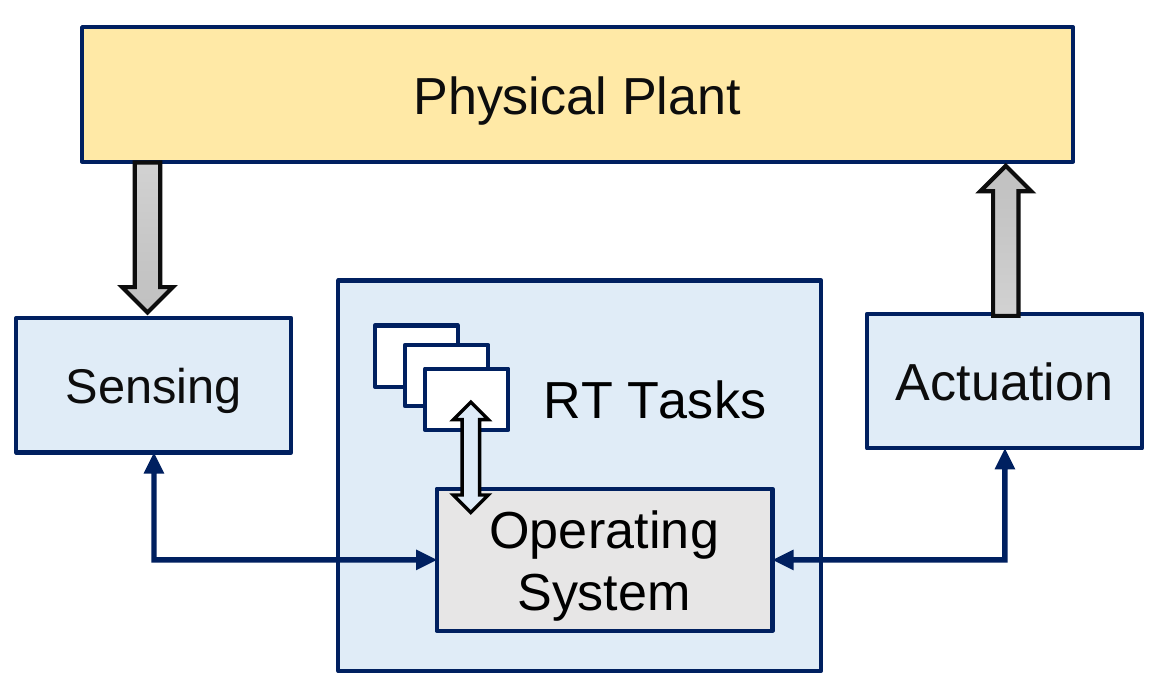}
    \caption{High-level schematic of a RT control system.} 
    \label{fig:controlsys}
\end{figure}

In Fig.~\ref{fig:controlsys} we present a high-level illustration of a RT control system. We consider a set of periodic RT control tasks $\Gamma = \{ \tau_i, \tau_2, \cdots, \tau_N\}$ that execute on single processor\footnote{Since majority of the RT-IoT edge devices still use single core chips due to simplicity and determinism.}. The physical system consists of a set of $M$ actuators (\eg servo, motor, buzzer): $\{ \pi_1, \pi_2, \cdots,  \pi_M \}$.  RT tasks periodically issue commands to the actuators to control physical entities (\eg wheel, propeller, alarm, robotic grip, \etc). We assume that each task is allowed to access a subset of peripherals. We represent this access permission as an $N \times M$ Boolean matrix $\mathbf{A} = [a_{ik}]$ where $a_{ik} = 1$ represents task $\tau_i$ can send commands to actuator $\pi_k$. We also assume that the RT tasks finish computation before their deadline, \eg the tasks are schedulable\footnote{In the Appendix
%~\ref{sec:wcrt_rt_tz} 
we present formal expressions to determine schedulability of the tasks.}. %Since the tasks are schedulable, the worst-case response time (WCRT) $R_i$ for each task $\tau_i \in \Gamma$ is less than its deadline\footnote{In the Appendix
%~\ref{sec:wcrt_rt_tz} we formally derive expressions to calculate WCRT.}.

\begin{figure*}[t]
    \centering
    % \hspace*{-2em}
    \vspace{-5em}
    \includegraphics[width=5.7in]{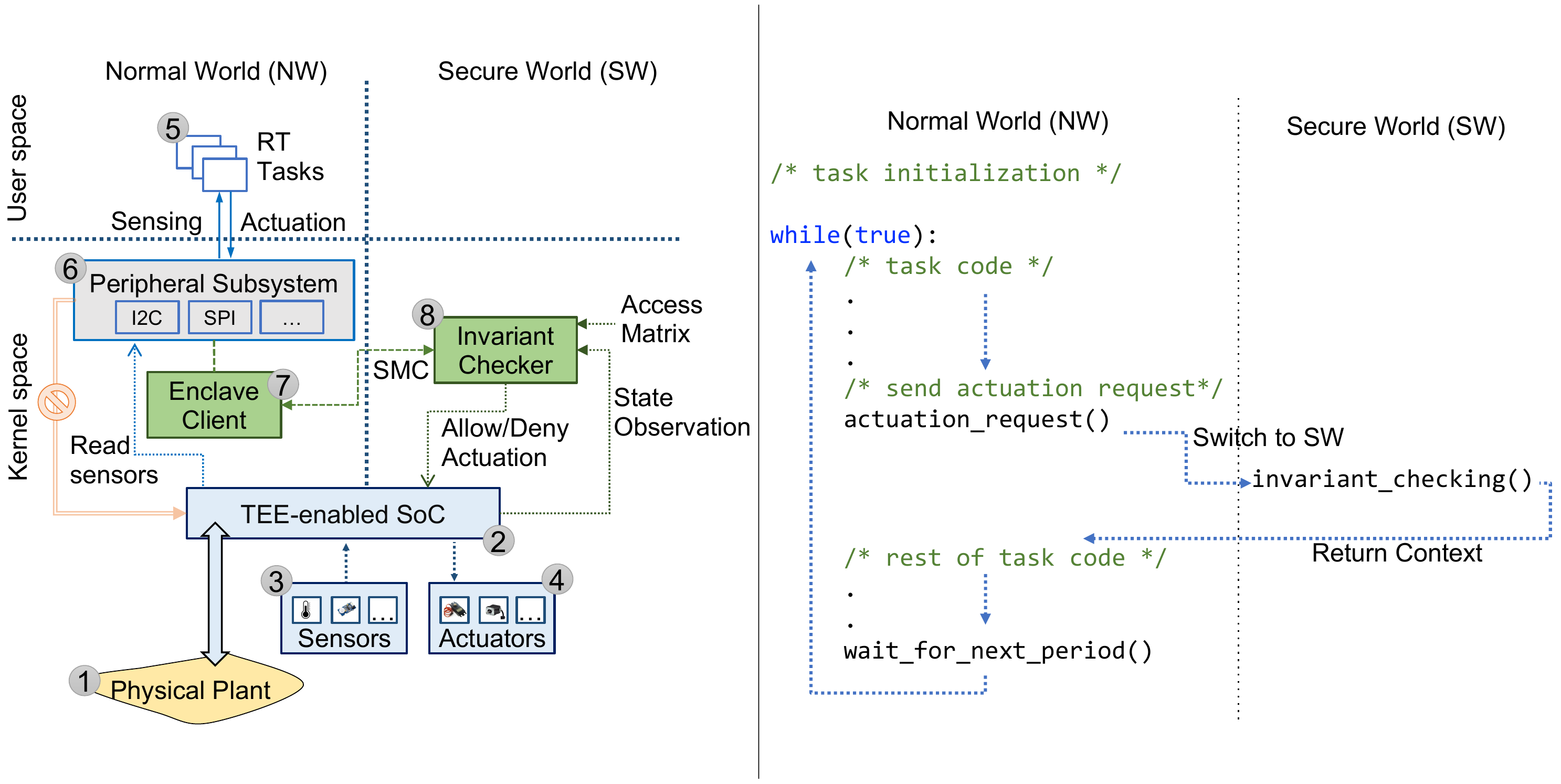}
    \caption{Overview of \pnametz system design (left) and high-level control flow of RT tasks in \pnametz (right).}
    \label{fig:sys_overview}
\end{figure*}

\subsection{Adversary Model} \label{sec:adv_model}

%Our assumptions on adversarial capabilities are similar to that considered in prior work~\cite{mhasan_rtss16, mhasan_ecrts17}.
We consider the following adversarial capabilities: 
\ca \textit{Integrity Violation} -- an adversary may insert a
malicious task (that respects the RT guarantees) and/or modify exiting control logic to manipulate actuator commands and control system behavior in undesirable
ways;
\cb \textit{Denial of Service (DoS)} -- the attacker may take control
of the RT task(s) and destabilize the physical plant \eg by sending multiple control requests in a burst that may  result in a malfunctioning actuator, or worse, damage the actual hardware/actuator and even threaten the safety of the system. %These threats are valid concerns as researchers demonstrated that such intrusions (to manipulate the critical systems) is not impossible~\cite{?}.

%We assume that the attacker has the capability to modify the data/contents being transferred to the actuators. This is a realistic assumption and the threat is a valid concern as demonstrated in prior work~\cite{?}. 

The attacker can gain privileged (\eg root) access to perform adversarial actions (\eg to spoof control signals). 
%We also (reasonably) assume that attacker wants to remain stealthy, \ie the adversary is likely to make small changes to the tasks rather than large-scale changes such as replacing the entire task (that may violate safety/timing constraints) to avoid immediate detection.   
We do
not make any assumptions as to how the compromised tasks enter the device. For instance, bad software engineering practices leave vulnerabilities in the systems~\cite{loi2017systematically}. When the system is developed using a multi-vendor model~\cite{sg2} (where its components are manufactured and integrated by different vendors) a malicious code logic may be injected (say by a less-trusted vendor) during deployment. The adversary may also induce end-users to download the
modified source code, say by using
social engineering tactics~\cite{securecore_syscal}. We also assume that the attackers do not have any physical access (\eg they can not physically control/turn off/damage the actuators).

% \begin{figure*}[t]
%     \centering
%     % \hspace*{-2em}
%     \vspace{-5em}
%     \includegraphics[width=5.65in]{Figures/tz_overview_cf.pdf}
%     \caption{Overview of \pnametz system design (left) and high-level control flow of RT tasks in \pnametz (right).}
%     \label{fig:sys_overview}
% \end{figure*}

\section{Actuation Monitoring Framework} \label{sec:act_mon_frm}

In the following we first introduce the \pnametz framework (\S\ref{sec:design}). We then present mechanisms to detect any abnormal control commands issued by (rogue) tasks and analyze schedulability conditions that ensures our (invariant) checking techniques can be enforced at runtime (\S\ref{sec:inv_n_timing}).

% \begin{figure}[t]
%     \centering
%     \includegraphics[width=2.5in]{Figures/tz_overview.pdf}
%     \caption{Overview of \pnametz System Design.}
%     \label{fig:sys_overview}
% \end{figure}

% \begin{figure}%[b]%[ht]
% 	\centering
% 	\hspace*{-3em}
% 	\begin{subfigure}[b]{0.4\linewidth} 
% 	    \includegraphics[width=2.2in]{Figures/tz_overview.pdf}
% 	  \caption{}
%     \label{fig:sys_overview}
% 	\end{subfigure}%
% 	\hfill
% 	\\
% 	\begin{subfigure}[b]{0.4\linewidth}
% 	    \includegraphics[width=2.1in]{Figures/tz_cf.pdf}
%     \label{fig:sys_cf}
%     \caption{}
% 	\end{subfigure}%
% 	\caption{\ca Overview of \pnametz System Design \cb Control flow of RT taks in \pnametz.}
% \end{figure}

\subsection{Overview and Architecture} \label{sec:design}

As mentioned earlier, to secure RT-IoT platforms we propose a TEE-based architecture
  that monitors actuation commands send to the physical entities. At the high-level, our design is based on the Simplex architecture~\cite{sha2001using}. 
 Researchers use Simplex-based architecture for time-critical cyber-physical systems to provide  fault-tolerance~\cite{liu2008ortega, l1_simplex} and recently, security~\cite{mohan_s3a, mhasan_resecure16, securecore}. A Simplex system consists of the following main components: \ca under normal operating conditions a \textit{High-Performance (complex)
Controller} actuates the physical plant (such a controller may be unverifiable due to its complexity, yet it must actuate a safety-critical system); \cb if, during operation the system state becomes unstable (\eg it is in
danger of violating a safety condition), a \textit{Safety
Controller} takes over and \cc the exact switching behavior is implemented by a \textit{Decision Module} that decides which controller output will drive the plant. In our context, we use a trusted (and verified) computing module (this is analogous to the safety controller) executed in a secure enclave (\viz \sw) and ensures  that even if the (potentially untrusted) \nw RT tasks are compromised, an adversary can not send false signals to the physical actuators.

 In Fig.~\ref{fig:sys_overview} we illustrate the high-level overview of \pnametz design and control flow of the RT tasks.  \pnametz contains the following essential components: \ca a \textit{TEE-enabled SoC} (System-on-Chip) such as those supported by ARM TrustZone~\cite{trustzone_survey} (block \textcircled{2} in the figure); \cb an \textit{Enclave Client} (block \textcircled{7}) that is used to communicate between \nw and \sw and \cc an \textit{Invariant Checker} (block \textcircled{8}) that is used to monitor (and validate) the actuation commands. The physical plant (\textcircled{1}) is connected with sensors (\textcircled{2}) and actuators (\textcircled{3}) and controlled by the (potentially vulnerable) RT tasks (\textcircled{5}). RT tasks execute in untrusted \nw and issue system calls (\eg \texttt{read()}, \texttt{write()}, \texttt{ioctl()}) to access the sensors/actuators using specific interface such as I2C~\cite{i2c} and/or SPI~\cite{spi}. \pnametz ensures that RT tasks cannot directly send any actuation commands (\eg it breaks the bridge between \textcircled{6}, \textcircled{2} and \textcircled{4}). We do this by placing a dispatcher (\eg enclave client) between the peripheral subsystem and actual hardware. As a result, before issuing any command to the physical actuators, it  will be validated by our trusted application (\eg invariant checker) running inside the secure enclave (\ie in the \sw). In particular, when a RT task $\tau_i$ sends an actuation command $x_{ik}^t$ to any peripheral $\pi_k$ at time $t$, enclave client traps those request and forwards the command to the   invariant checker using SMC. Depending on the access permission matrix $\mathbf{A}$ and current system state $\mathcal{S}(t)$, invariant checker then decides whether the given command $x_{ik}^t$ can be issued to the actuator $\pi_k$ (refer to \S\ref{sec:inv_n_timing} for details). In \pnametz, both the enclave client and invariant checker operate in the privileged mode (\eg kernel space) so that it can directly control low-level hardware. By using the enclave client (to invoke context switching) and invariant checking mechanisms, \pnametz ensures that even if the \nw RT tasks are compromised, an adversary can not send false signals to the actuators. We note that unlike \nw RT tasks that may perform other computation, the invariant checker contains a small, verified, code blocks that is used to monitor only actuation requests. We also note that \pnametz \textit{does not require any application-level modifications}, \eg developers can execute unmodified, existing legacy RT tasks, using our \pnametz enabled OS-kernel (refer to \S\ref{sec:implementaion} for implementation/porting details).

\begin{table*}[t b] 
\footnotesize
\centering
\caption{Applicability of \pnametz for Various RT-IoT Platforms}
\label{tab:rt_iot_exp}
\begin{tabular}{P{2.1cm} | P{1.7cm} |  P{1.7cm} |  P{4.95cm} | P{1.2cm} | P{4.1cm}} \hline 
\bf Platform & \bf Application Domain & \bf Actuators & \bf Possible Invariant Conditions\textsuperscript{*} & \bf Response & \bf Remarks \\
\hline \hline
Water/air monitoring system & Home/industrial automation & Buzzer, display & \ca Send high pulse to buzzer only if water-level is high/air quality abnormal/detect smoke; \cb do not display alert if the system state is normal & \tzignore & Ignore all commands that fail invariant checking \\ \hline
Surveillance system & Home/industrial automation  & Servo, buzzer & \ca Trigger alarm only if there is an impact/object detected in camera; \cb rotate camera (using servos) only within allowable pan/tilt angle & \tzignore & Ignore all commands that fail invariant checking \\ \hline
Infusion/syringe pump & Health-care & Motor, display & \ca Drive the motor only to allowable positions/rates \cb display only the amount of fluid infused (\eg obtained from motor encoders) & \tzignore & Ignore actuation when the task tries to infuse wrong amount of fluid \\ \hline
Robotic arm &  Manufacturing & Servo,  buzzer & \ca Check the servo pulse sequences matches with the desired (design-time) sequence; \cb do not raise alarm if the pulse sequence is normal & \tzignore, \tzfailsafe & If mismatch, use the predefined sequence; ignore other pulses using rate-control rule \\ \hline
Robotic vehicle (aerial/ground) & Manufacturing, surveillance, agriculture & Servo, motor & \ca Check if the robot is following the mission; \cb allow only predefined number of actuation commands per period & \tzignore, \tzfailsafe & Ignore command using rate-control rule. If it deviates from the mission, use predefined command and/or state-observations \\
\hline
\end{tabular}
\begin{flushleft}
\hspace*{1em}
\textsuperscript{*}We omit mathematical expressions for readability.
\end{flushleft}
\end{table*}

\subsection{Invariant Checking and Timing Analysis} \label{sec:inv_n_timing}

\subsubsection{Invariant Checking:} \label{sec:inv_check}

In order to validate each actuation command invoked by the RT tasks, \pnametz performs various actions. One obvious access control mechanism is to ensure that a task $\tau_i$ can access a given actuator $\pi_k$ only if the task has the required permission (\eg $a_{ik} = 1$). \pnametz therefore denies all the actuation commands from  tasks if the corresponding access flag is zero. However, if the attacker can compromise a task with legitimate access (to a given set of actuators) then the (victim) task may send arbitrary commands to the actuators. Therefore in addition to checking access matrix $\mathbf{A}$, \pnametz also performs checking of system invariants and monitors the number of actuation commands for a given time interval as we discuss below. 

\paragraph{State Invariant Checking:}
Invariant checking~\cite{adepu2017design} is useful to detect control spoofing attacks. For a given RT-IoT platform we do this by considering the availability of an invariant checking function $\mathsf{CheckInv}(\tau_i, \pi_k)$ that predicts the actuation signal and only allows access if the output of the function matches that of the requested command. In particular, if a task $\tau_i$ sends actuation command $x_{ik}^t$ at time $t$ to any peripheral $\pi_k$ and the task has the required permission (\ie $a_{ik} = 1$), $\mathsf{CheckInv}(\tau_i, \pi_k)$ first obtains system state  $\mathcal{S}(t)$ by observing a set of signals $S_i = \{s_1, s_2, \cdots, s_{L_i}\}$ and decides whether $x_{ik}^t$ is valid for current state $\mathcal{S}(t)$. 
For example, consider a warehouse water monitoring system where an alarm is triggered only if the water level of the tank (measured by the sensor $s_{WL}$) is higher that a predefined threshold ($\theta_{WL}$) and/or the water temperature ($s_{WT}$) is not in expected range (\ie $[\theta_{WT}^{t_1}, \theta_{WT}^{t_2}]$). We represent this as the following invariant rule: $\mathsf{INV}_W :: (s_{WL} > \theta_{WL})~\vee~(s_{WT} \notin [\theta_{WT}^{t_1}, \theta_{WT}^{t_2}]) \rightarrow x = \mathsf{ON} : x= \mathsf{OFF}$, \eg  \pnametz will only allow the sending of the high pulse (\ie $x = \mathsf{ON}$) to the alarm system (say a buzzer) only if the invariant conditions are satisfied. We note that since \pnametz operates at the kernel-level, it can directly access raw signals without any interaction of \nw RT tasks or other (user space) libraries. 

%For deeply embedded RT-IoT systems that do not preform any sensing, our $\mathsf{CheckInv}(\tau_i, \pi_k)$  function compares $j$-th actuation signal from $\tau_i$  (denoted by $x_{ik}^j$) by comparing  with the known actuation sequence to $\pi_k$, \ie $\Phi_{ik} = \{ \phi_{ik}^1, \phi_{ik}^2, \cdots,  \phi_{ik}^{J_{ik}}\}$ (\eg that stores in the \nw) and allow access to the actuator only if $x_{ik}^j = \phi_{ik}^j$. An example of such a system is the robotic arm in a manufacturing assembly line that periodically performs series of actuation commands (\eg pick $\rightarrow$ move $\rightarrow$ drop) to the objects.

\paragraph{Rate Control:}

Note that since RT systems are deterministic by design, the (worst-case) number of actuation requests can be bounded at design time~\cite{wcrt_survey}. Therefore, if a task $\tau_i$ tries to access actuator(s) more than expected within a given time interval (\eg $T_i$), it may be indication of a possible attack. In such cases \pnametz will limit subsequent access requests from $\tau_i$ and prevent the sending of actuation commands to the hardware. We enforce rate control using the following invariant rule: $\mathsf{INV}_{ik}^{RC} :: \Delta_{ik}(w) < \widehat{\theta}_{ik} \rightarrow \mathsf{CheckInv}(\tau_i, \pi_k) : \mathsf{ignore}$, \ie  \pnametz ignores further actuation commands if the number of requests $\Delta_{ik}(\cdot)$ from any job of $\tau_i$ within the (relative) time window $w \in [0, T_i]$ is exceeded a design-time threshold $\widehat{\theta}_{ik}$. Such a rate control mechanism is specially useful to defend against DoS attacks where an attacker sends multiple actuation commands in a burst (say to quickly change the speed of wheels/propellers in robotic ground/aerial vehicles, abruptly move robotic arms, falsely toggles buzzers, \etc) to disrupt normal operations of the system.

\subsubsection{Response Mechanisms:} \label{sec:inv_response}

When there exists a mismatch between output of the $\mathsf{CheckInv}(\cdot)$  and the requested actuation commands, \pnametz makes use of the following strategies to keep the physical system safe.

$\bullet$~~\tzignore: this strategy prevents the execution of any actuation commands requested by RT tasks. Hence, actuators will not receive any signals from \pnametz and will continue to operate using the last known (uncompromised) commands. \pnametz will also ignore commands if the task makes multiple requests in a short time window (\eg by using rate control rule).

$\bullet$~\tzfailsafe: while the \tzignore strategy ensures that actuators will not get any abnormal signals, ignoring actuation commands (for a long time)  may not be acceptable for highly dynamic systems such as unmanned ground/aerial vehicles (\eg it may crash). Therefore,  \pnametz also allows operation of a \tzfailsafe mode, \ie if it finds any mismatch, it ignores the requests from RT tasks and sends the predetermined (and/or based on the output of $\mathsf{CheckInv}(\cdot)$) commands to make the system safe/operational. As an example, if there is a sudden change in the propeller speed of a UAV, the \tzfailsafe strategy sets a safe, predefined speed, based on the current state of the UAV.

Depending on the target system, both of the above strategies may be required to keep the physical system operational. 
We note that invariant checking and response mechanisms are application dependent. \pnametz provides flexibility for the system engineers to develop appropriate mechanisms depending on the application requirements. In Table \ref{tab:rt_iot_exp} we summarize possible invariant conditions and response mechanisms that are applicable for various RT-IoT platforms -- however, this is by no stretch meant to an exhaustive list.

\subsubsection{Schedulability Analysis:} \label{sec:timing_analysis}

In order to perform invariant checking and execute the response mechanisms at runtime, we need to ensure that our framework should not cause delays and the timing requirements of RT tasks are satisfied (\eg they complete execution before deadline). We therefore develop design-time schedulability tests that ensure the taskset is schedulable (refer to the Appendix for details). For instance, the RT task $\tau_i$ is schedulable in \pnametz if the Worst Case Response Time (WCRT) $R_i^{TEE}$ is less than deadline, \ie $R_i^{TEE} = C_i^{TEE} + I_i^{TEE} \leq D_i$, where $C_i^{TEE}$ is the task WCET (including the time for world switching and invariant checking) and $I_i^{TEE}$ is the interference\footnote{In RT scheduling theory, the term `interference' refers to the amount of time (from release to deadline) the task $\tau_i$ is ready but can not be scheduled due to execution of other tasks.} from other tasks. The taskset $\Gamma$ is referred to as schedulable if all the tasks are schedulable, \viz $R_i^{TEE} \leq D_i, \forall \tau_i \in \Gamma$.

\section{Evaluation}

In this section we first present the implementation details of \pnametz (\S\ref{sec:implementaion}) and then show the viability of our approach using a case-study on a robotic vehicle (\S\ref{sec:rov_case_study}). Table \ref{tab:tz_param} summarizes the system configurations and implementation details.
%are summarized 

\begin{table}%[!thb]
\small
\caption{Summary of the Implementation Platform}
\label{tab:tz_param}
\centering
%\begin{tabular}{c||p{5.9cm}}
\begin{tabular}{P{2.5cm}||P{4.4cm}}
\hline %\\
\bfseries Artifact & \bfseries Configuration\\
\hline\hline
Platform         & Broadcom BCM2837 (Raspberry Pi 3)  \\
CPU         & 1.2 GHz 64-bit ARM Cortex-A53  \\
Memory & 1 Gigabyte  \\ %\hline
Operating System & Linux (\nw), OP-TEE (\sw)							\\ %\hline
Kernel version   & Linux kernel 4.16.56, \\ 
&                 OP-TEE core 3.4  \\ %\hline
Peripheral interface & I2C \\
Boot parameters &  \texttt{dtparam=i2c\_arm=on}, \\
                    & \texttt{dtparam=spi=on} \\ \hline
\end{tabular}
\end{table}

\subsection{System Implementation} \label{sec:implementaion}

We implemented a proof-of-concept prototype of
\pnametz on Raspberry Pi 3 (RPi3) Model B~\cite{rpi3} (equipped with 1.2 GHz 64-bit ARMv8 CPU and 1 GB RAM). We selected RPi3 as our implementation platform since \ca it  supports ARM
TrustZone and \cb previous research has shown feasibility of deploying
multiple IoT-specific applications on RPi3~\cite{securecore_syscal, cheng2017orpheus, virtsense_liu2018, protc_liu2017, liu2018alidrone}. We developed \pnametz using the Open-Portable
Trusted Execution Environment (OP-TEE)~\cite{optee} software stack that uses GlobalPlatform TEE APIs~\cite{globalplatform_tee_api} to provide TrustZone functionality. OP-TEE provides a minimal secure kernel (called OP-TEE core)
that can be run in parallel with the \nw OS (\eg Linux). In particular, we used  Ubuntu 18.04 filesystem with a  64-bit Linux kernel (version 4.16.56) as the \nw OS and our invariant checker is running on OP-TEE secure kernel (version 3.4). The enclave client was statically built with the Linux kernel. In order to implement the enclave client, we extended the Linux TEE interface (\texttt{/linux/drivers/tee/}) and enabled SMC from Linux kernel space\footnote{Since GlobalPlatform APIs only support SMC from user space.}.  We implemented the invariant checker as an OP-TEE kernel-level trusted application\footnote{This is known as PTA (Pseudo Trusted Application) in OP-TEE terminology.} (\eg in \texttt{/optee\_os/core/arch/arm/pta/}). In our current implementation \pnametz supports actuators that are controlled via the I2C interface. Specifically,  we modified the built-in structure \texttt{i2cdev\_fops} (\eg in \texttt{/linux/drivers/i2c/i2c-dev.c}) with our enclave client functions that is then switch the control to  the invariant checker (\eg by using  SMC). Our implementation code is available in a public repository~\cite{contego_tee_impl}.

\subsection{Case-Study: Robotic Vehicle} \label{sec:rov_case_study}

We implemented \pnametz in a COTS rover (named GoPiGo2, manufactured by Dexter Industries~\cite{gpg2}) that can be used in multiple IoT-specific applications such as remote surveillance, agriculture, manufacturing, \etc~\cite{guo2018roboads}.
%In our experiments we used GoPiGo2 robot manufactured by Dexter Industries~\cite{gpg2}. 
The rover is equipped with two optical encoders that are connected to the motors (\eg actuator in this setup): it can turn left by switching off the right encoder and vice-versa. The detailed specifications of the rover are available on the vendor website~\cite{gpg2}.

%\footnote{\url{https://www.dexterindustries.com}.}.
% \subsubsection{Platform Overview and Experimental Setup:}

\subsubsection{Results}

We first demonstrate how \pnametz can be used to protect such systems from actuation attacks and then measure the performance overheads.

\paragraph{Security Analysis:} 

For the following experiments, we conducted a line following mission where
the robot steered from an initial location to a target location
by following a line. The controller task was running as a \nw Linux application and executed vendor-provided  PID (Proportional–Integral–Derivative)
closed-loop control~\cite{gpg2_lf} to track the planned path using the data received from sensors. The rover used the following commands: $\mathsf{fwd()}$, $\mathsf{lft()}$, $\mathsf{rht()}$, $\mathsf{st\_sp}(\delta)$ for navigating the rover forward/left/right and set the speed to $\delta$, respectively, where each command sent a 5-byte value to the actuator registers (\eg wheel encoders/motors) using the I2C interface.  For this mission we defined the following three invariant conditions\footnote{We manually inspected the vendor-provided control code and translated them into invariant conditions.} that were used to monitor control signals (\eg $\mathsf{cmd}$):
$\mathsf{INV}_1 :: s_{LF} < -\theta \rightarrow \mathsf{cmd} = \mathsf{st\_sp}(\delta_1) \wedge \mathsf{rht()}$, 
$\mathsf{INV}_2 :: s_{LF} > \theta \rightarrow \mathsf{cmd}= \mathsf{st\_sp}(\delta_1) \wedge \mathsf{lft()}$ and
$\mathsf{INV}_3 :: s_{LF} \in  [-\theta, \theta] \rightarrow \mathsf{cmd}=\mathsf{st\_sp}(\delta_2) \wedge \mathsf{fwd()}$
%$\mathsf{INV}_4 :: s_{LF} \in  [-\theta, \theta] \rightarrow \mathsf{cmd=st\_sp(\delta_2) \wedge fwd()}$
where $s_{LF}$ was the readings from the sensor, $\theta=2500$ was a vendor-provided threshold (\eg to follow the line) and $\delta_1, \delta_2 \in [0, 255]$ were used to set the speed of the rover. We show the case where the access flag is set (\eg $a_{ik}=1$) since \pnametz will trivially deny requests if the corresponding flag is zero. In our experiments we used both the \tzfailsafe and \tzignore (to enforce rate control) strategies. For each actuation signal, our invariant checker matches with the desired signal and choose the appropriate strategy as we present in the following.

% \begin{figure}%[b]%[ht]
% 	\centering
% 	\begin{subfigure}[b]{0.5\linewidth} 
% 	    \hspace*{-1.5em}
% 		\centering \includegraphics[scale=0.285]{pgp_dos_pdis.pdf}
% 		\caption{\label{fig:gpg_dos_pdis}}
% 	\end{subfigure}%
% % 	\\
% 	\begin{subfigure}[b]{0.5\linewidth}
% 	    \hspace*{0.6em}
% 		\centering\includegraphics[scale=0.285]{Figures/pgp_dos_penb.pdf}
% 		\caption{\label{fig:gpg_dos_penb}}
% 	\end{subfigure}%
% 	\caption{TODO}
% \end{figure}

% \begin{figure}%[b]%[ht]
% 	\centering
% 	\begin{subfigure}[b]{0.5\linewidth} 
% 	    \hspace*{-1.5em}
% 		\centering \includegraphics[scale=0.285]{pgp_inv_pdis.pdf}
% 		\caption{\label{fig:gpg_inv_pdis}}
% 	\end{subfigure}%
% % 	\\
% 	\begin{subfigure}[b]{0.5\linewidth}
% 	    \hspace*{0.6em}
% 		\centering\includegraphics[scale=0.285]{Figures/pgp_inv_penb.pdf}
% 		\caption{\label{fig:gpg_inv_penb}}
% 	\end{subfigure}%
% 	\caption{TODO}
% \end{figure}

\begin{figure}%[b]%[ht]
	\centering
	\begin{subfigure}[b]{0.5\linewidth} 
	    \hspace*{-1.0em}
	    \centering\includegraphics[scale=0.285]{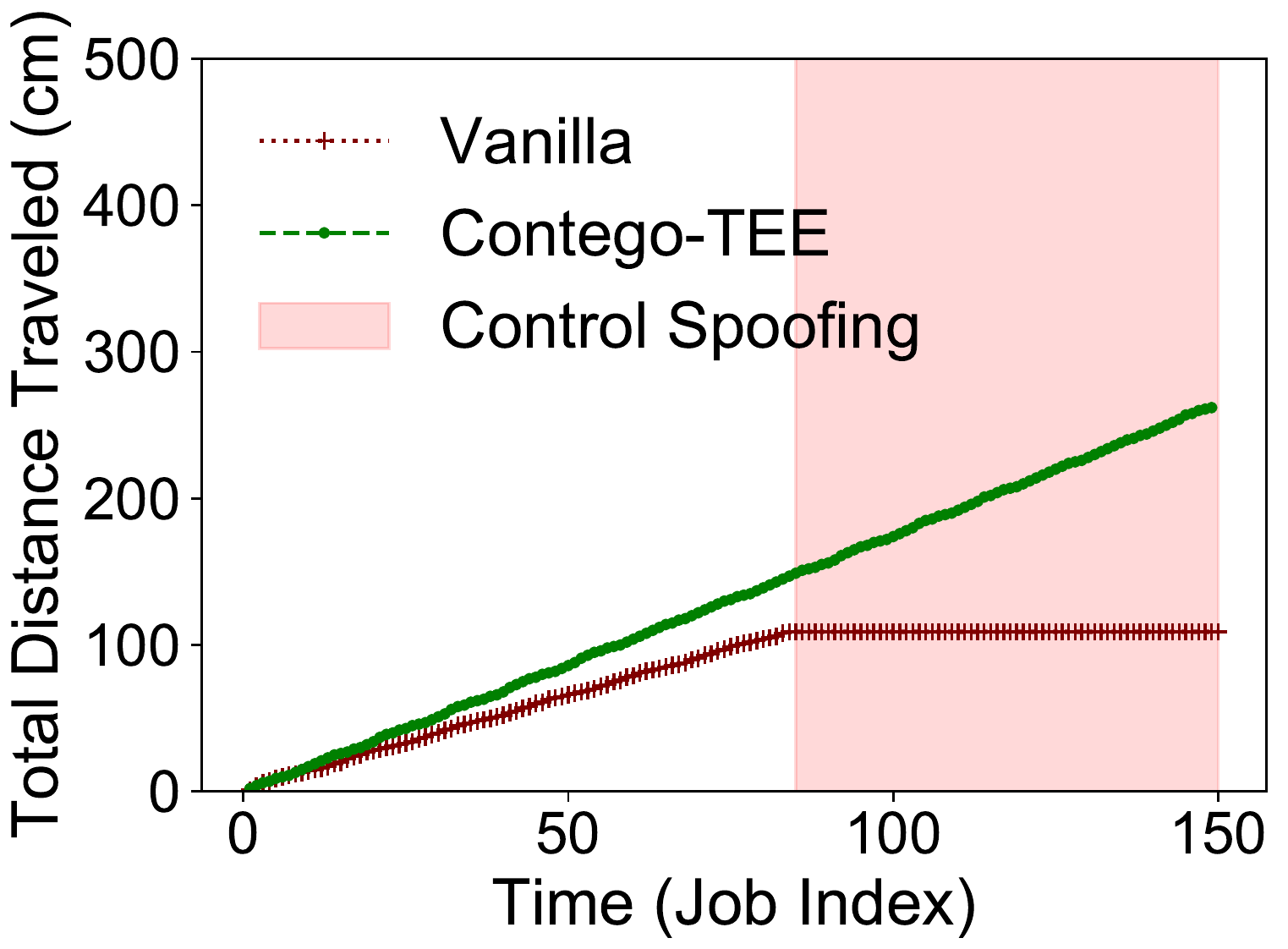}
		\caption{\label{fig:gpg_inv_both}}
		
	\end{subfigure}%
% 	\\
	\begin{subfigure}[b]{0.5\linewidth}
	   % \hspace*{0.6em}
	    \centering\includegraphics[scale=0.285]{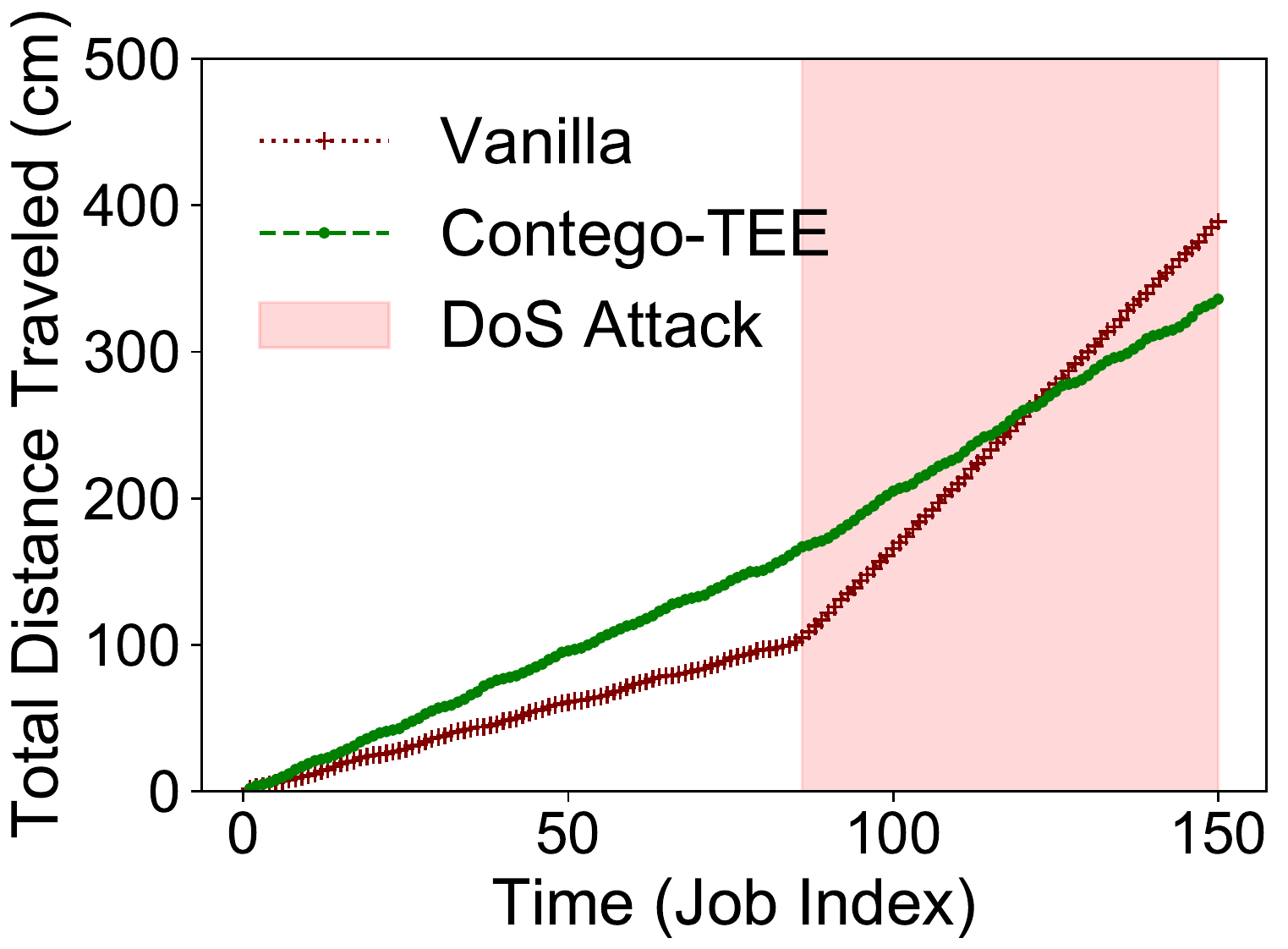}
		\caption{\label{fig:gpg_dos_both}}
	\end{subfigure}%
	\caption{Illustration of \pnametz under \ca control spoofing and \cb DoS attacks. \pnametz prevents the  sending of malicious commands to the motors and ensures that the rover moves at a steady speed.}
\end{figure}

In Fig.~\ref{fig:gpg_inv_both} we illustrate our invariant checking mechanism with \tzfailsafe strategy. The x-axis of the figure shows the time (\eg count of the controller job) and the y-axis is the total distance travelled by the rover (\eg readings from the encoders). In order to demonstrate malicious behavior, we followed a strategy similar to that considered in prior work~\cite{choi2018detecting, securecore, guo2018roboads, mhasan_resecure16}. In particular, during program execution, we injected a logic bomb (during the shaded region in  Fig.~\ref{fig:gpg_inv_both}) and sent erroneous commands to the controller. In this case, during the control spoofing attack, the rover deviated from the mission (\eg PID control loop) and falsely sent commands to turn off one of the motors. As a result, when \pnametz was not active, the rover was not following the line and the encoder readings (\ie traversed distance) remained same (see the maroon line in the figure).  We next executed the same code with \pnametz enabled (green curve in the figure). In this case, when each control command was issued, our checker followed the invariant conditions (\eg $\mathsf{INV}_i , 1 \leq i \leq 3$) and sent desired commands to the motors (and hence the rover was moving as expected).

We next show the effect of our rate control mechanism (Fig.~\ref{fig:gpg_dos_both}). In this experiment, when the DoS logic bomb was triggered (shaded region in the figure)
%although it followed the mission (\eg invariant conditions are satisfied), instead 
it sent multiple requests to increase the speed of the rover.  When \pnametz was not enabled, this caused the rover to move faster and hence there was a rapid increase in the encoder readings (\eg maroon line, shaded region in the figure).  In contrast, when \pnametz was active (green line), it disallowed multiple increase speed requests  per period (\eg according to \tzignore strategy) and hence the rover followed the line with a  steady speed.

\begin{figure}%[b]%[ht]
	\centering
	\begin{subfigure}[b]{0.5\linewidth} 
	    \hspace*{-1.0em}
		\centering \includegraphics[scale=0.285]{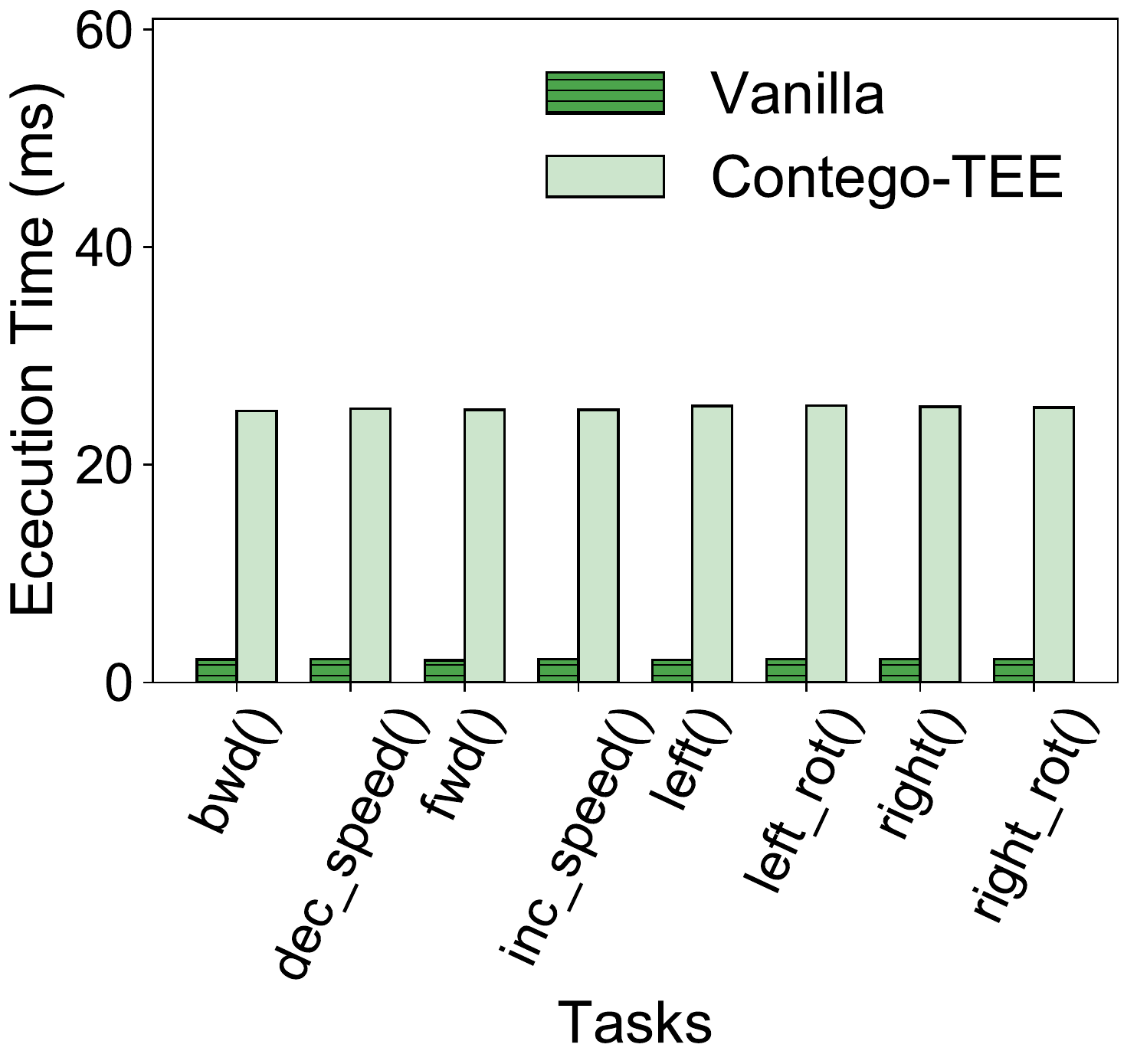}
		\caption{\label{fig:gpg_t99p}}
	\end{subfigure}%
% 	\\
	\begin{subfigure}[b]{0.5\linewidth}
	   % \hspace*{0.6em}
		\centering\includegraphics[scale=0.285]{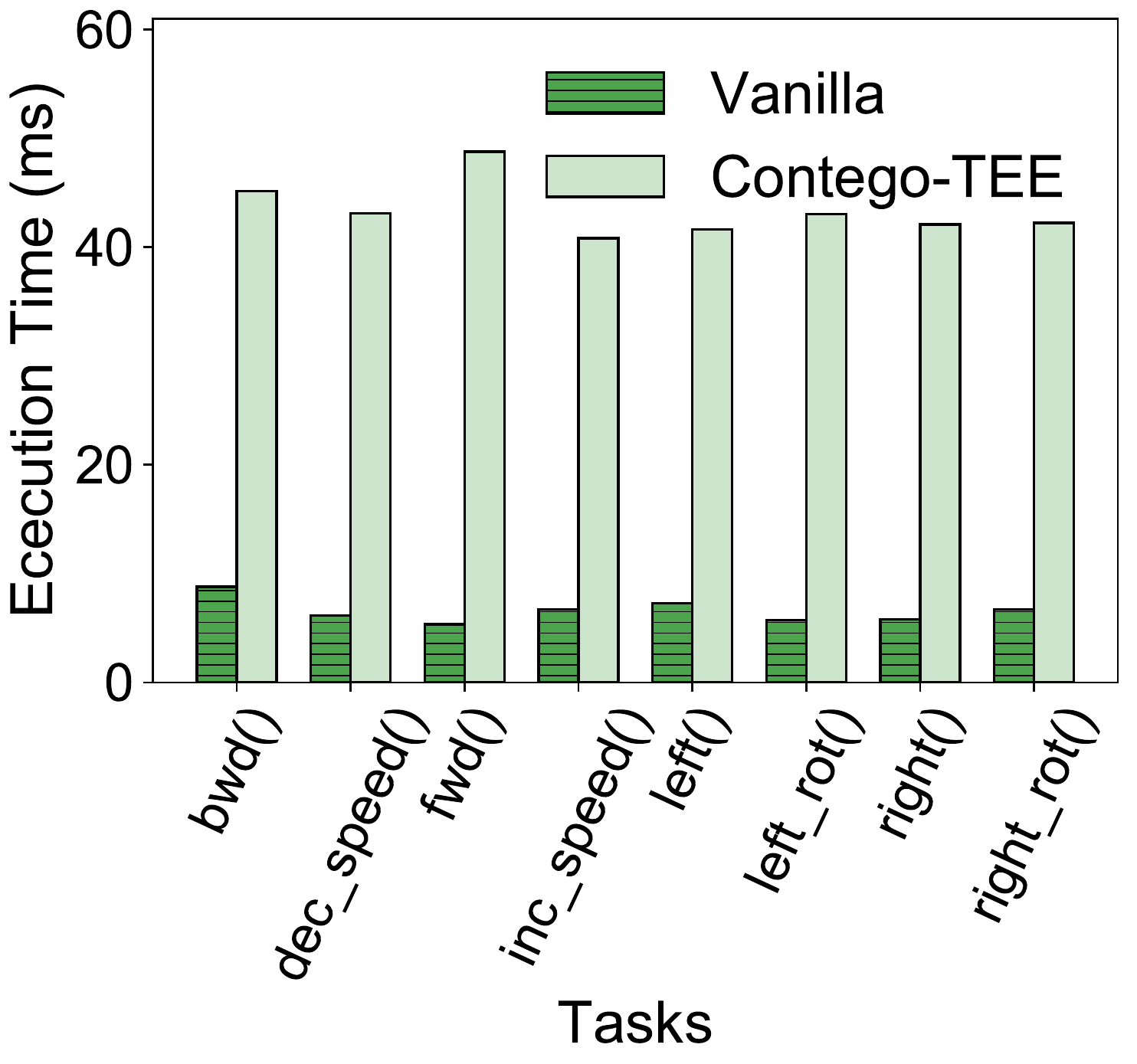}
		\caption{\label{fig:gpg_twcet}}
	\end{subfigure}%
	\caption{Runtime of rover control tasks with and without \pnametz: \ca for 99-th percentile and \cb worst-case. \pnametz increases the execution time by upto 43.47 ms (worst-case) and 23.31 ms (99th-percentile).}
	\label{fig:tz_overhead}
\end{figure}

\paragraph{Overhead Analysis:}

To measure the runtime overheads we conducted experiments with the vendor-provided control tasks~\cite{gpg2} as a benchmark  (Fig.~\ref{fig:tz_overhead}). In this setup our invariant checker was following a rate control policy and ignored more that one actuation request per period ($200$ ms). The x-axis of Fig.~\ref{fig:tz_overhead} shows the control tasks and y-axis represents execution time \ca when \pnametz is not enabled (dark bar) and \cb with \pnametz enabled (light bar). We present the timing results for 99th percentile (Fig.~\ref{fig:gpg_t99p}) and worst-case (Fig.~\ref{fig:gpg_twcet}). The timing values were measured using the Linux \texttt{clock\_gettime()} system call with \texttt{CLOCK\_MONOTONIC} clock parameter and we present data from $10,000$ trials. As we see from the figure, \pnametz increases the execution time -- this is expected due to (world) context switching  as well for invariant checking. From our experiments we found that \pnametz increases execution times by \ci 
34.11 to 43.47 ms (worst-case), \cii  
22.87 to 23.31 ms (99-th percentile) and \ciii 
19.55 to 19.60 ms (average-case) for the various control tasks and hence can be used with $15$ Hz (or slower) controllers (for this setup). This extra overhead results in increased security %(see next) 
and we expect this could be acceptable for various RT-IoT platforms.

\section{Related Work}

Enhancing security in time-critical cyber-physical systems is an active research area (see the related survey~\cite{mhasan_rtiot_sensors19}). Perhaps the closest line of work to ours is PROTC~\cite{protc_liu2017} where a monitor in the \sw
enforces secure access control policy (given by the control center) for
some peripherals of the drone and ensures
that only authorized applications can access
certain peripherals. Unlike our scheme, PROTC is limited for specific applications (\eg aerial robotic vehicles) and requires a centralized control center to validate/enforce security policies. In early work we proposed mechanisms to secure legacy time-critical systems~\cite{mhasan_ecrts17, mhasan_rtss16, mhasan_date18}. Researchers also proposed anomaly detection approaches for robotic vehicles~\cite{guo2018roboads, choi2018detecting, fei2018cross}. However these (prior) approaches do not provide any response mechanism and are vulnerable if the adversary can compromise the host OS. 

There exist various hardware/software-based mechanisms and architectural frameworks~\cite{mohan_s3a, securecore, securecore_memory, securecore_syscal, mhasan_resecure16, mhasan_resecure_iccps} to secure RT-IoT systems. However those frameworks are not designed to protect against control-specific attacks and may not be suitable for systems developed with COTS components. There also exist large number of research for generic IoT systems as well as use of TrustZone to secure traditional embedded/mobile applications (too many to enumerate here, refer to  the related surveys~\cite{yang2017survey, ammar2018internet, trustzone_survey, trustzone_survey_2}) -- however the consideration of time-critical and control-centric aspects of RT-IoT applications distinguish \pnametz from other research.
\section{Conclusion}

In this paper we presented a new framework named \pnametz that enhances the
security and safety of the RT-IoT systems. We use a combination of trusted hardware, intrinsic real-time nature and domain-specific characteristics of such systems to detect control intrusions and prevent the
physical plants from being misbehaved under attacks.   We believe our framework is tangential and can be incorporated into multiple RT-IoT and cyber-physical domains. 

%\newpage
\bibliographystyle{IEEEtran}

\bibliography{references_short}

\begin{thebibliography}{10}
\providecommand{\url}[1]{#1}
\csname url@samestyle\endcsname
\providecommand{\newblock}{\relax}
\providecommand{\bibinfo}[2]{#2}
\providecommand{\BIBentrySTDinterwordspacing}{\spaceskip=0pt\relax}
\providecommand{\BIBentryALTinterwordstretchfactor}{4}
\providecommand{\BIBentryALTinterwordspacing}{\spaceskip=\fontdimen2\font plus
\BIBentryALTinterwordstretchfactor\fontdimen3\font minus
  \fontdimen4\font\relax}
\providecommand{\BIBforeignlanguage}[2]{{%
\expandafter\ifx\csname l@#1\endcsname\relax
\typeout{** WARNING: IEEEtran.bst: No hyphenation pattern has been}%
\typeout{** loaded for the language `#1'. Using the pattern for}%
\typeout{** the default language instead.}%
\else
\language=\csname l@#1\endcsname
\fi
#2}}
\providecommand{\BIBdecl}{\relax}
\BIBdecl

\bibitem{mhasan_rtiot_sensors19}
C.-Y. Chen, M.~Hasan, and S.~Mohan, ``Securing real-time
  {Internet}-of-things,'' \emph{Sensors}, vol.~18, no.~12, 2018.

\bibitem{ddos_iot_camera}
J.~Westling, ``Future of the {Internet} of things in mission critical
  applications,'' 2016.

\bibitem{stuxnet}
N.~Falliere, L.~O. Murchu, and E.~Chien, ``W32. stuxnet dossier,'' \emph{White
  paper, Symantec Corp., Security Response}, vol.~5, p.~6, 2011.

\bibitem{security_medical}
S.~S. Clark and K.~Fu, ``Recent results in computer security for medical
  devices,'' in \emph{MobiHealth}, 2011, pp. 111--118.

\bibitem{checkoway2011comprehensive}
S.~Checkoway, D.~McCoy, B.~Kantor, D.~Anderson, H.~Shacham, S.~Savage,
  K.~Koscher, A.~Czeskis, F.~Roesner, T.~Kohno \emph{et~al.}, ``Comprehensive
  experimental analyses of automotive attack surfaces,'' in \emph{USENIX Sec.
  Symp.}, 2011.

\bibitem{i2c}
\BIBentryALTinterwordspacing
``{I}\textsuperscript{2}{C} manual,'' Philips Semiconductors, 2003. [Online].
  Available: \url{https://tinyurl.com/i2c-manual}
\BIBentrySTDinterwordspacing

\bibitem{rpi3}
``{Raspberry Pi},''
  \url{https://www.raspberrypi.org/products/raspberry-pi-3-model-b/}.

\bibitem{robot_arm_src}
``{Robot arm control},'' \url{https://github.com/tutRPi/6DOF-Robot-Arm}.

\bibitem{7345265_tee}
M.~{Sabt}, M.~{Achemlal}, and A.~{Bouabdallah}, ``Trusted execution
  environment: What it is, and what it is not,'' in \emph{IEEE
  Trustcom/BigDataSE/ISPA}, 2015, pp. 57--64.

\bibitem{trustzone_survey}
S.~Pinto and N.~Santos, ``Demystifying {ARM TrustZone}: A comprehensive
  survey,'' \emph{ACM CSUR}, vol.~51, no.~6, p. 130, 2019.

\bibitem{intel_sgx}
V.~Costan and S.~Devadas, ``{Intel SGX Explained},'' \emph{IACR Crypt. ePrint
  Arch.}, no. 086, pp. 1--118, 2016.

\bibitem{Liu_n_Layland1973}
C.~L. Liu and J.~W. Layland, ``Scheduling algorithms for multiprogramming in a
  hard-real-time environment,'' \emph{JACM}, vol.~20, no.~1, pp. 46--61, 1973.

\bibitem{res_time_rts}
N.~Audsley, A.~Burns, M.~Richardson, K.~Tindell, and A.~J. Wellings, ``Applying
  new scheduling theory to static priority pre-emptive scheduling,'' \emph{SE
  Journal}, vol.~8, no.~5, pp. 284--292, 1993.

\bibitem{bini2004schedulability}
E.~Bini and G.~C. Buttazzo, ``Schedulability analysis of periodic fixed
  priority systems,'' \emph{IEEE Trans. on Comp.}, vol.~53, no.~11, pp.
  1462--1473, 2004.

\bibitem{globalplatform_tee_api}
``{TEE client API specification v1.0},''
  \url{https://globalplatform.org/specs-library/tee-client-api-specification/}.

\bibitem{liu2018alidrone}
T.~Liu, A.~Hojjati, A.~Bates, and K.~Nahrstedt, ``Alidrone: Enabling
  trustworthy proof-of-alibi for commercial drone compliance,'' in \emph{IEEE
  ICDCS}, 2018, pp. 841--852.

\bibitem{loi2017systematically}
F.~Loi, A.~Sivanathan, H.~H. Gharakheili, A.~Radford, and V.~Sivaraman,
  ``{Systematically evaluating security and privacy for consumer IoT
  devices},'' in \emph{ACM IoTS\&P}, 2017, pp. 1--6.

\bibitem{sg2}
R.~Pellizzoni, N.~Paryab, M.-K. Yoon, S.~Bak, S.~Mohan, and R.~B. Bobba, ``A
  generalized model for preventing information leakage in hard real-time
  systems,'' in \emph{IEEE RTAS}, 2015, pp. 271--282.

\bibitem{securecore_syscal}
M.-K. Yoon, S.~Mohan, J.~Choi, M.~Christodorescu, and L.~Sha, ``Learning
  execution contexts from system call distribution for anomaly detection in
  smart embedded system,'' in \emph{ACM/IEEE IoTDI}, 2017, pp. 191--196.

\bibitem{sha2001using}
L.~Sha, ``Using simplicity to control complexity,'' \emph{IEEE Software},
  vol.~18, no.~4, pp. 20--28, 2001.

\bibitem{liu2008ortega}
X.~Liu, Q.~Wang, S.~Gopalakrishnan, W.~He, L.~Sha, H.~Ding, and K.~Lee,
  ``{ORTEGA: An efficient and flexible online fault tolerance architecture for
  real-time control systems},'' \emph{IEEE T. on Ind. Inf.}, vol.~4, no.~4, pp.
  213--224, 2008.

\bibitem{l1_simplex}
X.~{Wang}, N.~{Hovakimyan}, and L.~{Sha}, ``{L1Simplex: Fault-tolerant control
  of cyber-physical systems},'' in \emph{2013 ACM/IEEE ICCPS}, 2013, pp.
  41--50.

\bibitem{mohan_s3a}
S.~Mohan, S.~Bak, E.~Betti, H.~Yun, L.~Sha, and M.~Caccamo, ``{S3A}: Secure
  system simplex architecture for enhanced security and robustness of
  cyber-physical systems,'' in \emph{ACM international conference on High
  confidence networked systems}.\hskip 1em plus 0.5em minus 0.4em\relax ACM,
  2013, pp. 65--74.

\bibitem{mhasan_resecure16}
F.~Abdi, M.~Hasan, S.~Mohan, D.~Agarwal, and M.~Caccamo, ``{ReSecure}: A
  restart-based security protocol for tightly actuated hard real-time
  systems,'' in \emph{IEEE CERTS}, 2016, pp. 47--54.

\bibitem{securecore}
M.-K. Yoon, S.~Mohan, J.~Choi, J.-E. Kim, and L.~Sha, ``{SecureCore}: A
  multicore-based intrusion detection architecture for real-time embedded
  systems,'' in \emph{IEEE RTAS}, 2013, pp. 21--32.

\bibitem{spi}
\BIBentryALTinterwordspacing
``{SPI} block guide {V04.01},'' Motorola Inc, 2004. [Online]. Available:
  \url{https://tinyurl.com/spi-block}
\BIBentrySTDinterwordspacing

\bibitem{adepu2017design}
S.~Adepu and A.~Mathur, ``From design to invariants: Detecting attacks on cyber
  physical systems,'' in \emph{IEEE QRS-C}, 2017, pp. 533--540.

\bibitem{wcrt_survey}
R.~Wilhelm, J.~Engblom, A.~Ermedahl, N.~Holsti, S.~Thesing, D.~Whalley,
  G.~Bernat, C.~Ferdinand, R.~Heckmann, T.~Mitra \emph{et~al.}, ``The
  worst-case execution-time problem—overview of methods and survey of
  tools,'' \emph{ACM TECS}, vol.~7, no.~3, p.~36, 2008.

\bibitem{cheng2017orpheus}
L.~Cheng, K.~Tian, and D.~D. Yao, ``Orpheus: Enforcing cyber-physical execution
  semantics to defend against data-oriented attacks,'' in \emph{ACM ACSAC},
  2017, pp. 315--326.

\bibitem{virtsense_liu2018}
R.~Liu and M.~Srivastava, ``{VirtSense: Virtualize Sensing through ARM
  TrustZone on Internet-of-Things},'' in \emph{ACM SysTEX}, 2018, pp. 2--7.

\bibitem{protc_liu2017}
R.~\vspace*{0em}Liu and M.~Srivastava, ``{PROTC: PROTeCting drone's peripherals
  through ARM trustzone},'' in \emph{ACM DroNet}, 2017, pp. 1--6.

\bibitem{optee}
``{Open Portable Trusted Execution Environment},''
  \url{https://www.op-tee.org/}.

\bibitem{contego_tee_impl}
``{Implementation code for Contego-TEE},''
  \url{https://github.com/mnwrhsn/rt_actuator_security}.

\bibitem{gpg2}
``{GoPiGo},'' \url{https://github.com/DexterInd/GoPiGo}.

\bibitem{guo2018roboads}
P.~Guo, H.~Kim, N.~Virani, J.~Xu, M.~Zhu, and P.~Liu, ``{RoboADS: Anomaly
  detection against sensor and actuator misbehaviors in mobile robots},'' in
  \emph{IEEE/IFIP DSN}, 2018, pp. 574--585.

\bibitem{gpg2_lf}
``{Dexter Industries Sensors},'' \url{https://github.com/DexterInd/DI_Sensors}.

\bibitem{choi2018detecting}
H.~Choi, W.-C. Lee, Y.~Aafer, F.~Fei, Z.~Tu, X.~Zhang, D.~Xu, and X.~Xinyan,
  ``Detecting attacks against robotic vehicles: A control invariant approach,''
  in \emph{ACM CCS}, 2018, pp. 801--816.

\bibitem{mhasan_ecrts17}
M.~\vspace{0mm}Hasan, S.~Mohan, R.~Pellizzoni, and R.~B. Bobba, ``Contego: An
  adaptive framework for integrating security tasks in real-time systems,'' in
  \emph{Euromicro ECRTS}, 2017, pp. 23:1--23:22.

\bibitem{mhasan_rtss16}
M.~Hasan, S.~Mohan, R.~B. Bobba, and R.~Pellizzoni, ``Exploring opportunistic
  execution for integrating security into legacy hard real-time systems,'' in
  \emph{IEEE RTSS}, 2016, pp. 123--134.

\bibitem{mhasan_date18}
M.~Hasan, S.~Mohan, R.~Pellizzoni, and R.~B. Bobba, ``A design-space
  exploration for allocating security tasks in multicore real-time systems,''
  in \emph{DATE}, 2018, pp. 225--230.

\bibitem{fei2018cross}
F.~Fei, Z.~Tu, R.~Yu, T.~Kim, X.~Zhang, D.~Xu, and X.~Deng, ``Cross-layer
  retrofitting of {UAVs} against cyber-physical attacks,'' in \emph{IEEE ICRA},
  2018, pp. 550--557.

\bibitem{securecore_memory}
M.-K. Yoon, S.~Mohan, J.~Choi, and L.~Sha, ``Memory heat map: anomaly detection
  in real-time embedded systems using memory behavior,'' in \emph{ACM/EDAC/IEEE
  DAC}, 2015, pp. 1--6.

\bibitem{mhasan_resecure_iccps}
F.~Abdi, C.-Y. Chen, M.~Hasan, S.~Liu, S.~Mohan, and M.~Caccamo, ``Guaranteed
  physical security with restart-based design for cyber-physical systems,'' in
  \emph{ACM/IEEE ICCPS}, 2018, pp. 10--21.

\bibitem{yang2017survey}
Y.~Yang, L.~Wu, G.~Yin, L.~Li, and H.~Zhao, ``{A survey on security and privacy
  issues in Internet-of-Things},'' \emph{IEEE IoT J.}, vol.~4, no.~5, pp.
  1250--1258, 2017.

\bibitem{ammar2018internet}
M.~Ammar, G.~Russello, and B.~Crispo, ``{Internet of Things: A survey on the
  security of IoT frameworks},'' \emph{Elsevier J. of Inf. Sec. \& App.},
  vol.~38, pp. 8--27, 2018.

\bibitem{trustzone_survey_2}
W.~Li, H.~Chen, and H.~Chen, ``{Research on ARM TrustZone},'' \emph{ACM
  GetMobile}, vol.~22, no.~3, pp. 17--22, 2019.

\bibitem{joseph1986finding}
M.~Joseph and P.~Pandya, ``Finding response times in a real-time system,''
  \emph{The Comp. J.}, vol.~29, no.~5, pp. 390--395, 1986.

\end{thebibliography}

\appendix
\section*{APPENDIX}
\setcounter{section}{1}

\subsection*{Response Time Analysis for RT Tasks} \label{sec:wcrt_rt_tz}

% \subsection{Summary of Changes in the Linux Source Tree}

Our schedulability test is based on the fixed-priority response time analysis proposed in RT literature~\cite{res_time_rts}. Let $N_i$ be the number of actuation request for $\tau_i$ and $C_i^o$ is the additional computation time due to world switch and invariant checking. Then the WCET of $\tau_i$ can be represent as $C_i^{TEE} = C_i + N_i C_i^o$. Since our enclave client and invariant checker can serve one actuation request at a time (\eg an atomic process), $\tau_i$ may be delayed due to processing requests of lower priority tasks. Let $B_i^{TEE} = \max\limits_{\tau_j \in lp(\tau_i)} N_j C_j^o$ denote the `blocking' factor from tasks that are with lower-priority that $\tau_i$ (denoted as $lp(\tau_i)$). We note that the maximum computational demand for a given task $\tau_j$ in any interval length $0 \leq w \leq T_j$  can be no more than the maximum execution time required by one job of $\tau_j$ multiplied by the maximum number of jobs of $\tau_j$ that can execute in that interval~\cite{res_time_rts, bini2004schedulability}. The maximum interference experience by $\tau_i$ from other  tasks  for an interval $w$ can be expressed as: %$I_i^{TEE}(w) = 
$B_i^{TEE} + \sum\limits_{\tau_h \in hp(\tau_i)} \left\lceil \frac{w}{T_h} \right\rceil C_h^{TEE}$ where $hp(\tau_i)$ denotes the set of tasks with a  priority higher than $\tau_i$. Therefore, we can calculate the response time of $\tau_i$ (denoted as $r_i$) as follows:
\begin{equation}
    r_i = C_i^{TEE} + B_i^{TEE} + \sum\limits_{\tau_h \in hp(\tau_i)} \left\lceil \frac{r_i}{T_h} \right\rceil C_h^{TEE}.
\end{equation}
The WCRT then can be obtained by solving this recurrence using an iterative fixed-point search, \eg  $R_i^{TEE} = r_i^{(\alpha)} = r_i^{(\alpha-1)}$ for some iteration $\alpha$ with initial condition $r_i^{(0)} = 0$. The iteration is guaranteed to be converged if we assume that the total processor utilization (\ie $\sum\limits_{\tau_i}\tfrac{C_i^{TEE}}{T_i}$) is less than $1$~\cite{joseph1986finding}. The taskset is considered as `unschedulable' if there exists an $\alpha$ such that $r_i^{(\alpha)} > D_i$. Such unschedulability result will hint the designers to update parameters (\eg periods, number of tasks, invariant checking policies) to incorporate \pnametz framework for the target system.
% \section{Appendix 2}

\end{document}